# TREATING CANCER WITH STRONG MAGNETIC FIELDS AND ULTRASOUND

Friedwardt Winterberg, University of Nevada, Reno,
Email : winterbe@physics.unr.edu

## Abstract

It is proposed to treat cancer by the combination of a strong magnetic field with intense ultrasound. At the low electrical conductivity of tissue the magnetic field is not frozen into the tissue, and oscillates against the tissue which is brought into rapid oscillation by the ultrasound. As a result, a rapidly oscillating electric field is induced in the tissue, strong enough to disrupt cancer cell replication. Unlike radio frequency waves, which have been proposed for this purpose, ultrasound can be easily focused onto the regions to be treated. This method has the potential for the complete eradication of the tumor.

## 1. Introduction

The principal problem of any cancer treatment is the need for the almost complete destruction of all the cancer cells. This is so, because the "critical mass" from where a cancer can grow is not much less than 100 cells [1]. Even in surgery such a small number can easily escape through the bloodstream into other parts of the body, establishing there a metastatic colony. Chemotherapy has the potential to destroy all cancer cells, but because it is not a "magic bullet" attacking only cancer cells, it cannot be applied arbitrarily often. The same is true for radiation treatment.

The Achilles' heel of cancer cells is their increased vulnerability during mitosis. It is for this reason that I had in the past proposed to synchronize the cancer cell mitosis by "seeding" the cancer cells with small ferromagnetic particles, triggering their simultaneous mitosis with the suddenly applied magnetic field [2]. The thusly synchronized cancer cells can in this moment be exposed to a much higher dose of radiation, which integrated over time would be lower than what is possible without synchronization.

The "magic bullet" aspect of a successful cancer treatment is a call to physicists to search for such a "magic bullet".

## 2. Magnetic and Electric Field Treatment

Besides X-rays, the physicist has other means which conceivably might be used in treating cancer. They are:

1.  Magnetic fields.

2.  Electric fields.

3. Ultrasound.

All these fields are interesting because they can be externally applied and at critical intensities where they are harmful to cancer cells, but not harmful to the healthy cells.

Since cancer cells seem to have a smaller tensile strength than the normal cells, I had back in 1967 proposed to employ the strong magnetic fields, which can be produced with superconductors, for the inhibition of cancer cell growth [3]. It was, I believe, the first time that very strong magnetic fields were proposed in cancer therapy. Much more recently a research group in Israel has found that cancer cell replication can be disrupted in ~100 kHz alternating electric fields, with field strength of the order ~1 V/cm [4]. The same group also reports some in vivo evidence for the tumor growth inhibition in brain tumor patients.

## 3. **Combined Magnetic Field – Ultrasound Treatment**

In the reported ~100 kHz alternating electric field treatment, the wavelength of the electromagnetic radiation of the order of km, and cannot be focused onto a region as small as it is here required. At much higher frequencies focusing is possible but only with substantial heating. Also, because of the large impedance of cellular membranes, the electric field cannot easily diffuse into the interior of a cell.

To overcome these difficulties it is here proposed to generate the electric field by ultrasound inductively in situ. If placed in a strong magnetic field **B** the ultrasound makes the tissue oscillate against the magnetic field inducing in it the electric field

$$\mathbf{E} = (1 / c) \mathbf{v} \times \mathbf{B} \tag{1}$$

where **v** is the velocity of oscillation and c is the velocity of light. **E** is measured in electrostatic units esu with 1 esu = 300 V/cm.

Let us assume the tissue can withstand a sound wave pressure p = $10^6$dyn/cm$^2$ (one atmosphere), then the velocity of the oscillating tissue is ($\rho \approx$ 1g/cm$^3$):

$$v \simeq \sqrt{p / \rho} \simeq 10^3 \, \text{cm/s} \tag{2}$$

For B = 100 kG which can be reached with superconductors, one finds that E = 1V/cm. Instead of a superconductor one may use a ferromagnet. For gadolinium with a saturation field strength of 60 kG, the same electric field would be reached with a velocity v = 1.7 x $10^3$ cm/s, with a sound pressure $\rho v^2 \approx$ 3 atmospheres, probably still non-destructive.

We have to check if the magnetic field is not "frozen" into the oscillating tissue. The time $\tau$ for the magnetic field to penetrate the tissue of a thickness x is given by

$$\tau = \frac{4 \pi \sigma x^2}{c^2} \qquad (3)$$

where $\sigma \approx 10^{12}$ s$^{-1}$, is the electrical conductivity of blood. For an oscillating sound field in water or tissue where the velocity of sound is a few km/s, the wave length is of the order of a few cm.

Setting in (3) x~3cm, one obtains $\tau \sim 10^{-7}$ s, and therefore with $\omega \sim 10^5$ s$^{-1}$, $\omega\tau$ <<1. With regard to the oscillating tissue the magnetic field therefore remains static in the rest frame reference system.

With a wavelength of a few cm, the ultrasound wave can be easily focused down to a few cm. But because of the **v** x **B** dependence of the electric field, only the velocity component perpendicular to **B** leads to an induced electric field **E**. This suggests to focus and direct ultrasonic waves from at least 4 spatial directions onto the tumor (better still from 6 directions).

## 4. <u>Discussion</u>

In the experiments reported by the Israeli research team, the tumor disrupting effect was observed for a field of ~1V/cm. But because there the electric field comes from an electromagnetic wave which does not easily pass through a cellular membrane, in contrast to the inductively in situ generated electric field resulting from the time dependence of the magnetic vector potential which is not blocked by the cell membrane, an electric field somewhat less than 1V/cm may lead to the same effect. This then would make it possible to use magnetic fields produced by magnetic solenoids with ferromagnetic cores, where field strength of 20 kG is typical. For v = $10^3$ cm/s, it would lead to E ~ 0.2 V/cm. Furthermore, the electromagnets could there be run from the AC low frequency (60 cycles/second) power net.

By repeatedly scanning the tumor with the ultrasound wave focus, this concept may have the potential for an almost 100% destruction rate, without any damage to the healthy tissue, realizing the "magic bullet" idea.

## <u>Referenecs</u>